\documentclass[%
reprint,
 amsmath,amssymb,
 aps,
prb,
]{revtex4-1}
\usepackage{graphicx}
\usepackage{dcolumn}
\usepackage{bm}

\begin{document}

\preprint{APS/123-QED}

\title{Thermal expansion coefficient of single crystal silicon from 7 K to 293 K}

\author{Thomas Middelmann, Alexander Walkov, Guido Bartl and Ren\'{e} Sch{\"o}del}
\affiliation{Physikalisch-Technische Bundesanstalt, Bundesallee 100, 38116 Braunschweig, Germany}

\date{\today}

\begin{abstract}
We measured the absolute lengths of three single crystal silicon samples by means of an imaging Twyman-Green interferometer
in the temperature range from 7~K to 293~K with uncertainties of about 1~nm. 
From these measurements we extract the coefficient of thermal expansion with uncertainties in the order of $1\times 10^{-9} {\rm /K}$. 
To access the functional dependence of the length on the temperature usually polynomials are fitted to the data. 
Instead we use a 
physically motivated model equation with 7~fit parameters for the whole temperature range. 
The coefficient of thermal expansion is obtained from the derivative of the best fit. 
The measurements conducted in 2012 and 2014 demonstrate a high reproducibility and the agreement of two independently produced samples supports single crystal silicon as reference material for thermal expansion. 
Although the results for all three samples agree with each other and with measurements performed at other institutes, they significantly differ from the recommended values for thermal expansion of crystalline silicon.

\end{abstract}

\pacs{}
\maketitle

\section{\label{sec:intro}Introduction}

High accuracy knowledge of the coefficient of thermal expansion (CTE) of material samples is essential 
for development
and characterization of ultra stable materials needed e.g. in semiconductor industry, precision optics or in aerospace applications\cite{eny07,eny12}.
To enable high accuracy operation many measurement systems use reference material samples as calibration standards. For this purpose single crystal silicon (SCS) is a commonly used material that is particularly suited. 
SCS is available as off-the-shelf product with high purity. Together with its single crystal structure this guaranties supplier independent thermal expansion, that is isotropic and low compared to metals. Further it offers a high thermal conductivity, especially towards cryogenic temperatures, which supports a homogeneous temperature distribution over the sample.

Thermal expansion of ultra stable materials at cryogenic temperatures is of increasing scientific interest 
 driven by technological applications\cite{sch12,eny12,sto11a,sto10,kar06,bia06}.
High accuracy CTE values at cryogenic temperatures are essential e.g. for construction and operation of spaceborn telescopes as the Herschel Space Observatory\cite{pil10} (operated at 85~K),
the James Webb Space Telescope\cite{gar06} (to be operated below 50~K), 
or the SPICA telescope\cite{fer09} (desired: 5~K). In a round robin performed a few years ago under the guidance of the European Space Agency (ESA) and the Centre Spatial de Li$\rm\grave{e}$ge (CSL) the thermo-mechanical characterization abilities of some of the world-wide best institutes were compared \cite{roo11}. A clear lack of high accuracy measurement capabilities at cryogenic temperatures was pointed out. 
For the design of ultra stable structures it is necessary to characterize the thermal expansion of these materials with an uncertainty below $3\times 10^{-9} {\rm /K}$ in the entire in-service temperature range \cite{tec09}.
Thus, for a reference material as SCS a similar or even lower uncertainty is indispensable. To realize this PTB's Ultra Precision Interferometer (UPI) was equipped with an extended measurement pathway, enabling absolute length measurements from 7~K to 293~K\cite{sch12}.

A review of measurement techniques for the CTE of metals and alloys is given by James et al.\cite{jam01}. 
Often the CTE is determined from the ratio of length change over temperature change $\Delta l / \Delta T$ 
(as e.g. by Lyon et al.\cite{lyo77}) requiring additional corrections to account for the finite non-zero interval size \cite{kro77}. 
Instead of measuring length changes, we measure the absolute length as a function of temperature. 
Further, rather than fitting polynomials in several regions to the data, we use a physically motivated model equation, that covers the whole investigated temperature range, with only seven free parameters. 
By this we avoid an overestimation of the data, and give a coherent description of the functional dependence.

In this paper we report on CTE measurements on three different samples, manufactured from two independent suppliers. All measurements agree with each other, and with other high accuracy measurements. Albeit a comparison with the CTE values recommended for crystalline silicon unveils significant systematic deviations of 3 to 4~$\sigma$ in a wide temperature range (90~K to 210~K).

\section{\label{sec:exp}Experimental setup}
\begin{figure}[b]
\centering
\includegraphics[width=\columnwidth]{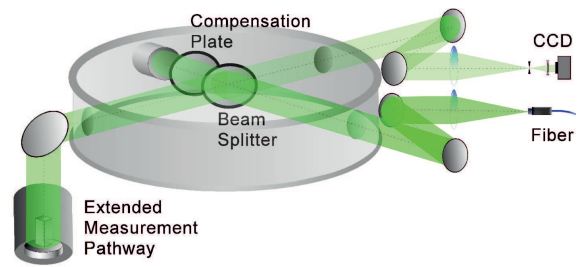}
\caption{Schematic of the beam path in PTB's Ultra Precision Interferometer.  Prismatic samples are placed in the extended probe beam on the left. Its temperature can be varied between 7~K and 293~K by a pulse tube cryocooler.}
\label{fig:setup}
\end{figure}

The experimental setup has been described in detail earlier\cite{sch12}. For completeness we recall the main aspects in short. 
The measurements were performed with PTB's UPI, an imaging Twyman-Green interferometer operated under vacuum (Fig.~\ref{fig:setup}). The expanded beam from three alternatively used frequency stabilized lasers is split up into a reference and a probe beam. The probe beam is directed towards the sample, and the arising interference pattern is imaged on a CCD-Array.
The length measurements are performed by phase stepping interferometry. For each length measurement 10 pictures of the fringeless interference pattern at 10 different phase steps are recorded. The length is determined twice, from 5 pictures for each value, and both values are averaged. Phase stepping is realized by tilting the compensation plate in the reference beam by a piezo stepper.

For measurements at cryogenic temperatures the probe beam is elongated by an extended measurement pathway (EMP),  housed  in a separated vacuum chamber. The cooling is achieved by a pulse tube cryocooler (PTC) with two cooling stages, enabling the variation of the sample temperature from 7~K to 293~K. The samples are surrounded by a copper shield and measurements are performed under a low pressure ($p\approx 1$~mbar) helium atmosphere that guaranties a homogeneous temperature distribution. 
The optical path dilatation due to the refractive index of diluted helium is regarded as described earlier\cite{sch12}. The temperature is measured by means of two Rhodium-Iron-Resistance-Thermometers (RIRT), which are fitted into holes in the center of the samples and dummy-samples respectively. The RIRT sensors have been calibrated by PTB and their measurement uncertainty rises from 15~mK at a few K to 25~mK at 293~K. 

At temperatures below 35~K the PTC is still operating during measurements. To minimize the influence of vibrations the camera is triggered by the frequency of the PTC. At temperatures above 35~K the PTC is switched off and measurements are performed while the temperature slowly drifts upwards. This restricts the size of temperature intervals that can be investigated within one day. 
But since the absolute length of the samples is measured, it is not necessary to perform all measurements in one run. 
Indeed within one cooling cycle an arbitrary chosen temperature interval of about 15~K was investigated. 
By this also the emergence of deposits on the surfaces of samples and base plate, that were observed at longer times ($>1$ day) at low temperatures, was prevented.

\section{\label{sec:meas}Measurements}

We have performed two measurement series: one in 2012 on sample~$\#1$    
and one in 2014 on samples~$\#2$    
and~$\#3$.    
Samples~$\#1$ and~$\#2$ have been prepared from SCS material which was acquired from Wacker Siltronic, Germany by PTB some years ago, as mentioned by Becker\cite{bec99} and Sch\"odel\cite{sch01}. 
These samples were cut in $\left\langle 100\right\rangle$ orientation from the 0-Zone  of  a high purity dislocation free float-zone(FZ)-silicon crystal. The level of impurities was found very low (oxygen: $(1 \ldots 2)\times 10^{15}\ {\rm cm^{-3}}$, carbon:  $(2 \ldots 6)\times 10^{15} \ {\rm cm^{-3}}$, nitrogen: $<10^{14} \  {\rm cm}^{-3}$) and the 0-Zone is free from extended swirl defects. Further information on the material can be found in\cite{bec99}. The gauge block shaped samples have a length of about 35~mm and two parallel faces of 9~mm~$\times$~20~mm cross section. 
The latter were lapped by Kolb \& Baumann GmbH \& Co. KG (KoBa) to optical quality. Sample~$\#3$ was acquired by PTB together with a ``Reference Material Certificate'', containing certified CTE values for this sample, from the National Metrology Institute of Japan (NMIJ)\cite{nmij}. This sample was as well made from a high-purity FZ-silicon crystal produced by Shin-Etsu Handotai Co., Ltd. This third sample is a rectangular block with a length of 30~mm and two parallel faces of 10~mm~$\times$~10~mm (Fig. 2). 
These faces have been lapped in the same way as for the other two samples at KoBa to optical quality. For the measurements the samples were wrung onto a base plate also made from SCS and polished in the same way by KoBa. 

In the 2012 measurement series not always optimal conditions have been used as was discovered later on. Reasons for suboptimal conditions were: Deposits from condensation due to the presence of residual air or small leakages in the vacuum system caused by temperature gradients. Measurements that were subject to deposits were suspended from the 2012 measurement series. Also a few measurements had to be suspended due to high local or temporal temperature gradients, that were caused by unilateral or too strong heating. 

In the 2014 measurement series high local or temporal temperature gradients were avoided and deposits from condensation were prevented by using a protective helium atmosphere during cool down and between measurements as well as a cold trap operated at 77~K with liquid nitrogen. 
The protective helium atmosphere first of all reduces the pressure difference to ambient air pressure and by this also leakage amounts. 
Second of all it dilutes residual gases and prevents their condensation. 
Additionally the pressure gradient between the EMP inner vacuum chamber and the surrounding isolating vacuum chamber is inverted, and leakages from the surrounding vacuum chamber are suppressed completely. For the pressure of the protective helium atmosphere a high pressure of 900~mbar turned out to be most effective. A side effect was the shorter time needed to cool down, due to better thermal linking by the heat transport of the protective atmosphere. In general the times at cryogenic temperatures were kept short, i.e. typically under 8~hours and the protective helium gas was exchanged about every three hours, to dilute contaminations.  

The temperature sensors described in Sec.~\ref{sec:exp} were placed in drillings in the center of sample~$\#1$ and ~$\#2$. Sample~$\#3$ is too small to contain a temperature sensor thus its temperature is determined as average of two sensors placed in two silicon blocks that were wrung to the same base plate (one of them is sample~$\#2$). 
Measurement uncertainties are discussed in detail in Appendix~\ref{sec:u}.

\section{\label{sec:cte}CTE determination}

The coefficient of thermal expansion is determined according to the ISO definition~\cite{iso}
\begin{equation}
\label{eq:alphaiso}
\alpha(T)=\frac{1}{l_{\rm RT}}\frac{dl(T)}{dT},
\end{equation}
with the temperature dependent length $l(T)$ and the length $l_{\rm RT}$ at room temperature ${\rm RT}=293.15$~K. 

To determine the CTE according to Equation~\ref{eq:alphaiso}, the functional dependence of the length on the temperature needs to be known. 
This can be achieved by fitting an appropriate function to the data. 
Thereto usually polynomials are used and fitted in several regions to the data. 
But of course selecting polynomial order and fitting regions also introduces uncertainties.
Thus to obtain a possibly low uncertainty, a functional description over the whole temperature range, that is derived from physics is desirable.

In a simple model, that neglects phonon dispersion, the CTE can be expressed according to the Gr{\"u}neisen equation~\cite{iba69}
\begin{equation}
\label{eq:alphagruen}
\alpha(T)=\frac{1}{3}\kappa\gamma c_v(T),
\end{equation}
as being proportional to the specific heat $c_v(T)$ per unit volume at constant volume.
Here also the temperature dependencies of the Gr{\"u}neisen parameter $\gamma$ and harmonic compressibility $\kappa$ are neglected.
A simple model for the temperature dependence of the specific heat is provided by the Einstein model of specific heat:
\begin{equation}
\label{eq:einstein}
c_v(T)\propto \left( \frac{\Theta_E}{T}\right)^2 \frac{e^{\theta_E/T}}{(-1+e^{\theta_E/T})^2}.
\end{equation}
Here $\Theta_E$ is the material dependent Einstein-Temperature. Equations~\ref{eq:alphagruen} and~\ref{eq:einstein} represent only an approximate functional dependence of the CTE on the temperature. But they can serve to constitute an appropriate fit function by summation over several so called ``Einstein-Terms''~\cite{iba69}, such that the CTE can be described by
\begin{equation}
\label{eq:alphafit}
\alpha(T)=\frac{1}{l_{\rm RT}} \sum\limits_{k=1}^{m} a_k \left( \frac{\Theta_k}{T}\right)^2 \frac{e^{\theta_k/T}}{(-1+e^{\theta_k/T})^2},
\end{equation}
and the fit function is obtained by integration to be
\begin{equation}
\label{eq:lfit}
l(T)=l_0+ \sum\limits_{k=1}^{m} a_k  \frac{\theta_k}{-1+e^{\theta_k/T}}.
\end{equation}
The fit parameters $\Theta_k$ have the unit of temperature but do not directly represent physical properties, although they are roughly related to peaks in the frequency spectrum of the phonon density of states. The fit parameter $l_0$ corresponds to the sample's length at T = 0 K. For the measurements considered here on SCS $m = 3$, i.e. $n = 7$ fit parameters are sufficient to fit the data very well within the expected uncertainty. 
Fitting and uncertainty evaluation of the CTE values are described in Appendix~\ref{sec:ucte}.

\section{\label{sec:results} results}

\begin{figure}[!t]
\centering
\includegraphics[width=\columnwidth]{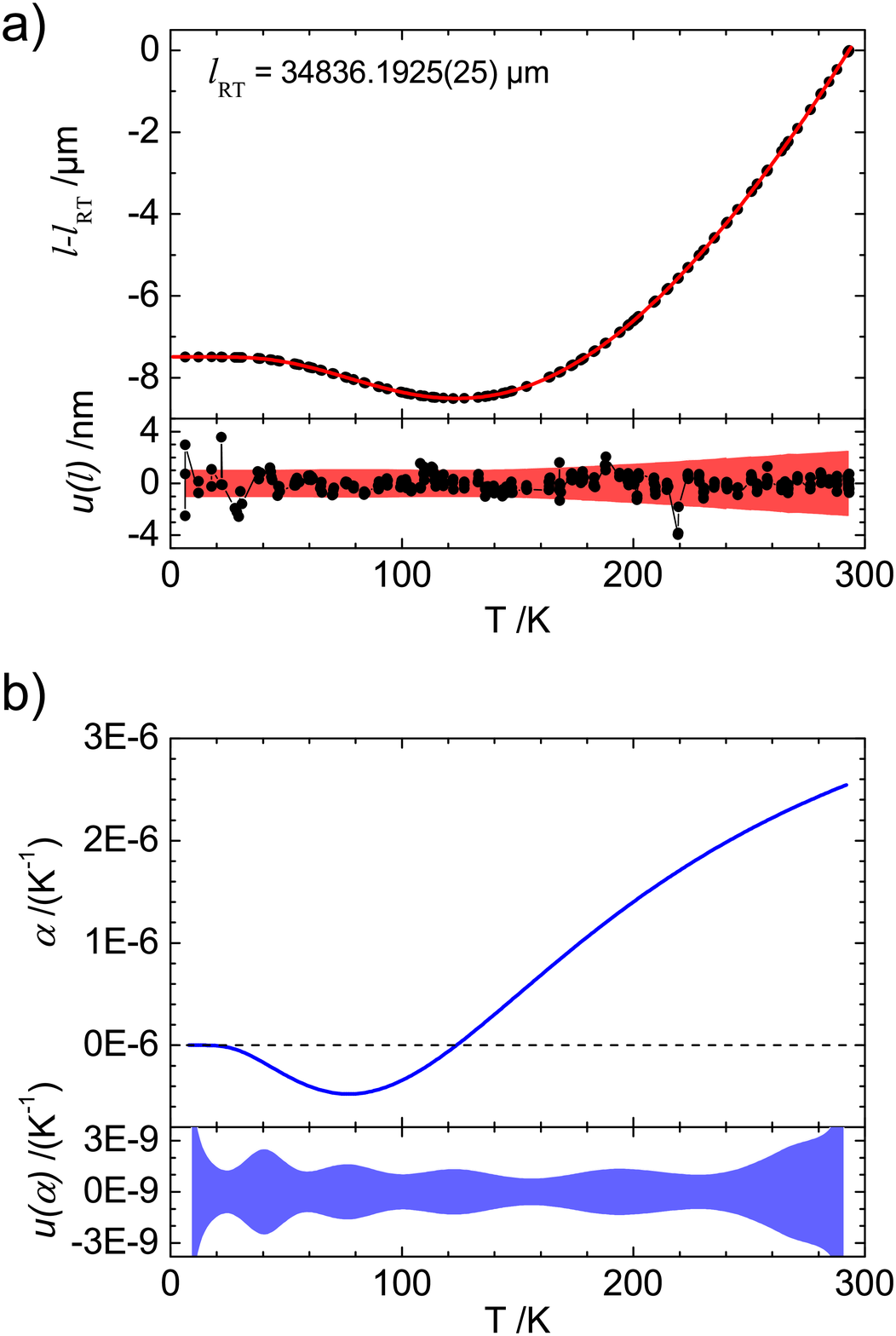}
\caption{a) Measured length versus temperature for sample~$\#2$, fit (red line) according to Eq.~\ref{eq:lfit}, and below residuals from the fit and overall uncertainty of the length versus temperature data (Appendix~\ref{sec:u}). b) Deduced CTE according to Eq.~\ref{eq:alphafit} and overall uncertainty (Appendix~\ref{sec:ucte}).}
\label{fig:fitres}
\end{figure}

The length versus temperature data obtained for sample~$\#2$ are plotted in Fig~\ref{fig:fitres}a
together with the fit (Equation~\ref{eq:lfit}) and its residuals. The latter are small compared to the overall uncertainty (red) estimated in Appendix~\ref{sec:u}, demonstrating the appropriateness of the fit function. The fit parameters are tabulated in Tab.~\ref{tab:param} in Appendix~\ref{sec:ucte} and the results obtained for sample~$\#2$ are tabulated in Tab.~\ref{tab:values} in Appendix~\ref{sec:values}

provided online as supplemental material.
The corresponding CTE values are plotted in Fig~\ref{fig:fitres}b (Equation~\ref{eq:alphafit}) and below its overall uncertainty (blue), whose estimation is described in Appendix~\ref{sec:ucte}. 

\begin{figure}[!t]
\centering
\includegraphics[width=\columnwidth]{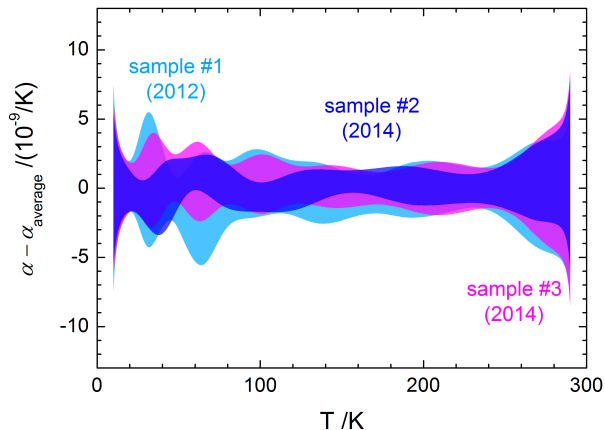}
\caption{Comparison of all three measurements performed at PTB: In 2012 on sample~$\#1$ (light blue), and in 2014 on 
samples~$\#2$ (dark blue) and~$\#3$ (purple). Plotted is a band with a half width of 1~$\sigma$ around the respective CTE minus the average CTE of all three measurements.}
\label{fig:ctevglptb}
\end{figure}

The results for all three samples
are compared in 
Fig~\ref{fig:ctevglptb}.
Plotted is the respective CTE as a band (half width of 1~$\sigma$), minus the average of all three measurements. 
All three measurements agree very well within their uncertainties. 
The largest difference of two of the measurements is smaller than $3\times 10^{-9} / {\rm K}$. 
The best results are obtained for sample~$\#2$, since it was investigated with the optimized measurement procedure. 
Furthermore in comparison to sample~$\#3$, that was investigated in the same measurement series, 
it contained a temperature sensor placed in its center, whereas the temperature of sample~$\#3$
was determined as average of two sensors placed in adjacent samples, as described in Sec.~\ref{sec:meas}.

\begin{figure}[!t]
\centering
\includegraphics[width=\columnwidth]{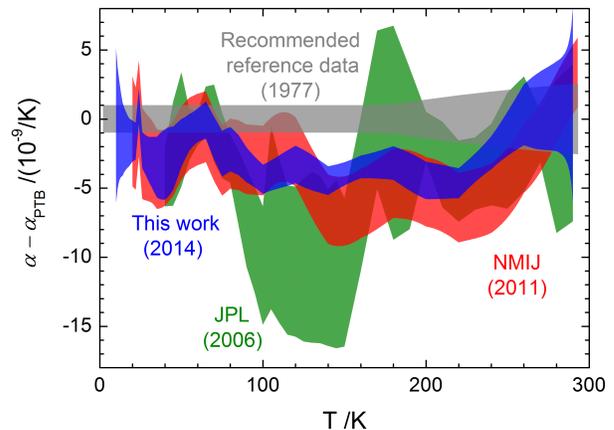}
\caption{Comparison of high accuracy CTE measurements performed on crystalline silicon. 
Plotted are bands with a half width of 1~$\sigma$ around the respective CTE minus recommended reference data\cite{swe83}. Compared are: The results from this work on sample~$\#2$ (blue), the data from the reference certificate acquired from NMIJ \cite{nmij} (red), measurements from JPL\cite{kar06} (green), and the 1-$\sigma$-uncertainty of the reference data (gray).}
\label{fig:ctevgllit}
\end{figure}

A comparison with other\cite{Note2} high accuracy CTE measurements performed on crystalline
silicon is presented in 
Fig~\ref{fig:ctevgllit}.
Plotted is a band with a half width of 1~$\sigma$ around the respective CTE, minus the CTE values recommended by CODATA as reference data\cite{swe85}. The results from this work (sample~$\#2$) are plotted in blue, the data from the reference certificate acquired from NMIJ \cite{nmij} in red and measurements from the Jet Propulsion Laboratory (JPL) in green\cite{kar06}. 
The 1-$\sigma$-uncertainty of the reference data is indicated by the gray shaded band according to the original publication \cite{swe83}, that is commonly cited. Although it is worth to be mentioned, that the data in the recommendation of CODATA \cite{swe85}, are provided with an expanded uncertainty of $2\times10^{-8}/{\rm K}$ for temperatures between 40~K to 300~K, which is up to 20 times larger than the uncertainty stated originally. 

All but the recommended reference data agree within their uncertainties in the whole temperature range. 
But in the temperature range from 90~K to 230~K the recommended reference data differ from the results of this work by about 
$4\times10^{-9}/{\rm K}$ or $3\sigma$ to $4\sigma$.
Here it is noticed, that the measurements \cite{lyo77} which are the basis for the reference data at temperatures below 300~K, were performed on a polycrystalline but not monocrystalline silicon sample, as in all other measurements presented here. 

The temperature measurements in this work refer to the ITS-90\cite{pre90}, while for the recommended reference data the IPTS-68 was used, as mentioned by Kroeger and Swenson\cite{kro77}. 
Taking this into account reduces the mismatch of both measurements slightly, but not significantly. 
The differences between thermodynamic temperatures and the ITS-90 were estimated by Fischer et al.\cite{fis10} and are smaller than 9~mK in the temperature range considered in this work, which is less than half the temperature uncertainty (Sec.~\ref{sec:exp}) and thus negligible for our results.

\begin{figure}[!t]
\centering
\includegraphics[width=\columnwidth]{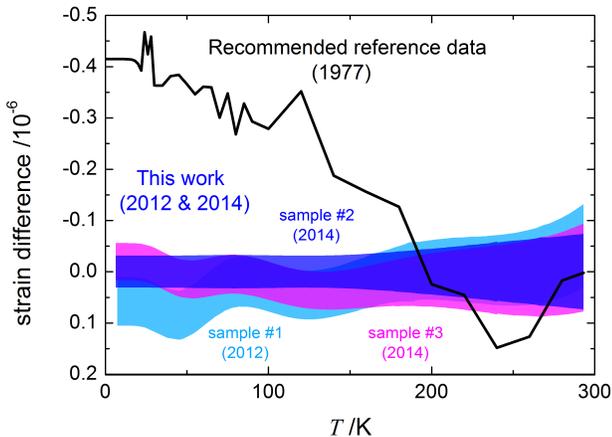}
\caption{Comparison of thermal strain results. 
Plotted are bands with a half width of 1~$\sigma$ around the respective strain minus the strain measured for sample~$\#2$.  Compared are: The results from this work on sample~$\#1$ (light blue), on sample~$\#2$ (dark blue) and on sample~$\#3$ (purple) with the recommended reference data~\cite{lyo77} (black line), which are provided without uncertainty.
}
\label{fig:straincomp}
\end{figure}

The significance of the mismatch between the recommended reference data and this work becomes more evident considering the thermal strain $s(T)=(l(T)-l_{\rm RT})/l_{\rm RT}$.  
In Figure~\ref{fig:straincomp}
the recommended thermal strain data are compared with the results from this work. 
While the strain difference of all three measurements from this work is compatible with the estimated uncertainties,
the strain difference between this work and the recommended data increases towards lower temperatures and becomes as large as $13~\sigma$, referring to the measurement on sample~$\#2$. As the strain is proportional to the integration of the CTE, the difference visible in Fig.~\ref{fig:ctevgllit} is now accumulated, emphasizing the significance of the discrepancy. This also illustrates the advantage of absolute length measurements, since the lengths measured at different temperatures can be directly compared and do not require an integration.

\section{\label{sec:conclusion}conclusion}

The CTE
of three SCS samples were determined from absolute length measurements at temperatures between 7~K and 293~K.
Uncertainties in the order of or smaller than $3\times10^{-9}/{\rm K}$ were obtained
(10~K to 290~K).
The two independent measurement series performed in 2012 and in 2014 demonstrate a high degree of coincidence, no matter if CTE or thermal stain are compared. Furthermore one of the two samples investigated in 2014 was manufactured in Japan, completely independent from the other two samples manufactured in Germany. Thus also the universality of SCS as reference material is demonstrated with an accuracy of a few times $10^{-9}/{\rm K}$. 

A comparison with other high accuracy measurements performed in the last decade demonstrates a good agreement. Albeit a mismatch is found with the reference data recommended by CODATA\cite{swe83,swe85} in a wide temperature range (90~K to 230~K). The CTE difference in this region corresponds to $4\times 10^{-9}/{\rm K}$, which is about four times the uncertainty. This mismatch is supported by the measurements at the other institutes. 
In recent years accuracies below $3\times 10^{-9}/{\rm K}$ are coming more and more to the fore in industrial applications, thus reliable reference data are mandatory. As SCS is commonly used as reference material for thermal expansion, also the recommended reference data should be based on SCS. 

%
%
%
\acknowledgments
We thank Stan Heltzel (ESA) and Michael Krystek (PTB) for helpful discussions.
This work was financially supported by ESA-project ``QT6986 Characterization of ultra-stable materials at cryogenic temperature''.

%
%
%
%
%

\appendix
\section{\label{sec:u}Measurement uncertainty}

The uncertainty of the individual data points stems from both, the length and the temperature measurements. In general each of these contributions consists of a type A (evaluated by statistical methods) and a type B (not evaluated by statistical methods) part \cite{gum08}.
The uncertainty of the pure length measurements is evaluated as described in \cite{sch12}. The type A uncertainty is estimated to be $u_{\rm A}(l)=1$~nm and the type B uncertainty to be $u_{{\rm B}}(l)=0.42$~nm.
In the case of sample~$\#2$ a settling of the wringing contact of $0.039\ {\rm nm/day}$ was observed, as apparent from length measurements performed at room temperature. The settling was corrected for.

The uncertainty of the temperature measurements is treated as type B uncertainty. 
It is estimated according to the GUM-rules\cite{gum08} for a rectangular distribution from the temperature difference of both sensors and the sensor uncertainty to be 
\begin{equation}
\label{eq:ut}
u_{\rm B}(T)=\sqrt{\frac{(T_2-T_1)^2}{12}+(0.016+3.3\cdot 10^{-5}\ T)^2}.
\end{equation}
The uncertainty of temperature $u(T)$ is propagated to the length dimension by
\begin{equation}
\label{eq:utprop}
u_{{\rm B},T}(l)=u_B(T) \sqrt{\left(\frac{dl}{dT}\right)^2+u\left(\frac{dl}{dT}\right)^2} ,
\end{equation}
with $u(\frac{dl}{dT})=2\times 10^{-9} l_{\rm RT}$ being the estimated uncertainty of the derivative to account for the uncertainty that remains when the derivative equals zero.

The total uncertainty of the length versus temperature is obtained from the sum of squares to be
\begin{equation}
\label{eq:ulall}
u(l)=\sqrt{ u_{\rm A}(l)^2 + u_{\rm B}(l)^2 + u_{{\rm B},T}(l)^2},
\end{equation}
and is plotted in Figure~\ref{fig:fitres}a for sample~$\#2$.

\section{\label{sec:ucte}CTE uncertainty}

Equation~\ref{eq:lfit} is fitted to the length versus temperature data
using the Levenberg-Marquardt algorithm with Mathematica 8.5 (Wolfram Research) by the built-in function NonlinearModelFit. For fitting, the data are weighted according to the combined measurement uncertainty from the individual length and temperature measurements with $1/u^2=1/(u_{\rm A}(l)^2+u_{{\rm B},l}(l)^2+u_{{\rm B},T}(l)^2$.
\begin{table}
\centering
\caption{Values of the fit parameters obtained for sample~$\#2$ and corresponding statistical uncertainties.}
\label{tab:param}
\begin{tabular}{ c c c c }
\hline
Parameter			&				&Value													&$u(p_i)$ 						\\
\hline
$p_1=a_1$			&	&$-3.398\times10^{-08}$							&$6.3\times10^{-11}$ 	\\  	
$p_2=\theta_1$&	&$199.61$		  											&$0.20$								\\
$p_3=a_2$ 		&	&$1.487\times10^{-07}$						&$3.2\times10^{-10}$	\\
$p_4=\theta_2$&	&$612.00$													&$0.61$								\\
$p_5=a_3$			&	&$3.496\times10^{-08}$							&$4.4\times10^{-10}$	\\
$p_6=\theta_3$&	&$890.05$													&$0.09$								\\
$p_7=l_0$			&	&$0.0348286997$											&$1.5\times10^{-10}$	\\
\hline
\end{tabular}
\end{table}

Some of the $n=7$ fit parameters $p_i$ (Tab.~\ref{tab:param}) are strongly correlated. Hence, to estimate the type A uncertainty of the fit the law of propagation of uncertainty, for correlated input quantities has to be applied \cite{gum08}.
This requires, to take into account both, the variances $c_{ii}=u(p_i)^2$, and also the covariances $c_{ij}$ of the parameters $p_i$ and $p_j$,
being the diagonal and the off-diagonal elements of the  $n\times n$ covariance matrix 
of the fit. 
Thus, the type A uncertainty of the fit is given by
\begin{equation}
\label{eq:ualfit}
u_{\rm A}(l(T))=\sqrt{\sum\limits_{i=1}^{n}\sum\limits_{j=1}^{n} \frac{\partial l}{\partial p_i} \frac{\partial l}{\partial p_j} c_{ij}},
\end{equation}
in which $\partial l/\partial p_i$ are the partial derivatives of the Fit (Eq.~\ref{eq:lfit}) with respect to the parameter $p_i$. 
Analogously the type A uncertainty of the CTE is given by 
\begin{equation}
u_{\rm A}(\alpha(T))=  \sqrt{\sum\limits_{i=1}^{n}\sum\limits_{j=1}^{n} \frac{\partial \alpha}{\partial p_i} \frac{\partial \alpha}{\partial p_j} c_{ij}},
\label{eq:uaalphafit}
\end{equation}
with $\partial \alpha/\partial p_i$ being the partial derivatives of Eq.~\ref{eq:alphafit} with respect to the parameter $p_i$. 

Additionally a type B error can occur due to inappropriateness of the fit function.  To analyze this we consider the adjacent median of the fit residuals. From its common mode amplitude we deduce an estimate of the type B error of $u_{\rm B,model} (l(T))=1$~nm. Since the GUM~\cite{gum08} does not offer a simple model for the propagation of the type B uncertainty of a quantity to its derivative, the type B uncertainty of the fitted length 
\begin{equation}
u_{\rm B}(l(T))=  \sqrt{u_{\rm B}(l)^2+u_{{\rm B},T}(l)^2+u_{\rm B,model} (l(T))^2}
\label{eq:ublfitted}
\end{equation}
is projected via the ratio of the type A uncertainties of the CTE and the length (equations~\ref{eq:uaalphafit} and~\ref{eq:ualfit}), such that the correlation is regarded here as well
\begin{equation}
u_{\rm B}(\alpha(T))= \frac{u_{\rm A}(\alpha(T))}{u_{\rm A}(l(T))}u_{\rm B}(l(T)).
\label{eq:ubalphafitted}
\end{equation}

At last we consider a circumstance inherently connected with the determination of the slope of measured data. 
The closer two data points are, the harder it is to tell the slope of a line connecting them. 
As the line could go from the lower end of the uncertainty interval of one point, to the upper end of the uncertainty interval of the other, and vice versa. Since the measured data are distributed over a wide temperature range, this becomes only crucial at the borders of the investigated temperature range. Especially for low temperatures, when T approaches zero, according to equation~\ref{eq:alphafit} applies $\lim\limits_{T\rightarrow 0} \alpha(T) = 0$. As a matter of principle the same applies for the type A uncertainty (equation~\ref{eq:uaalphafit}): $\lim\limits_{T\rightarrow 0} u_{\rm A} (\alpha) = 0$  and the type B uncertainty (equation~\ref{eq:ubalphafitted}), i.e. $\lim\limits_{T\rightarrow 0} u_{\rm B} (\alpha) = 0$.
This is a property of the fit function used, which is only an approximation and reality might differ. 
Although it can be derived from the third law of thermodynamics, that $\alpha(T=0)=0$, for small temperatures above $T=0$ the true shape might differ from the fit function. Here it is noticed, that in this temperature region small positive CTE values as large as $1\times 10^{-9}/{\rm K}$ at $T=14$~K  have been reported~\cite{swe83}. 
As this is in the order of the uncertainty, it cannot be validated by the measurements presented in this work.
We estimate this slope uncertainty to be
\begin{equation}
u_{\partial}(\alpha(T))=\frac{\sqrt{2} u_{\rm A}(l)}{l_{\rm RT}(\Delta T-2\left|\overline{T}-T \right|)},
\label{eq:uslope}
\end{equation}
here $u_{\rm A}(l)$ is the type A uncertainty of the individual length measurements, $\Delta T=T_{\rm max}-T_{\rm min}$ is the size of the temperature range under investigation and $\overline{T}=T_{\rm min}+\Delta T/2$ is the central temperature. 
The slope uncertainty $u_{\partial}(\alpha(T))$ is small in the largest part of the investigated temperature range, but becomes the dominating contribution below 10~K and above 290~K and avoiding an overestimation of the data in these regions.

All three contributions are combined to give the overall uncertainty of the CTE
\begin{equation}
u(\alpha)=\sqrt{u_{\rm A}(\alpha(T))^2+u_{\rm B}(\alpha(T))^2+u_{\partial}(\alpha(T))^2}, 
\label{eq:ualpha}
\end{equation}
that is plotted in Figure~\ref{fig:fitres}b for sample~$\#2$.

\section{\label{sec:values} Tabulated CTE values}

\begin{table}
\centering
\caption{Tabulated CTE values $\alpha\ /(10^{-9}/{\rm K})$ and uncertainties $u(\alpha)\ /(10^{-9}/{\rm K})$  versus temperature $T\ /$K, extracted from the measurements on sample~$\#2$.}
\label{tab:values}
\begin{tabular}{ r r r}
\hline
$\ \ \ \ \ \ \ \ \ \ T$						&$\ \ \ \ \ \ \ \ \ \ \ \ \alpha$&$\ \ \ \ \ \ u(\alpha)$ $ \ \ \ \ \  $	\\
\hline
8.15	&	0.0	&	11.2	$ \ \ \ \ \  $ \\
13.15	&	-0.1	&	3.0	$ \ \ \ \ \  $ \\
18.15	&	-2.0	&	1.7	$ \ \ \ \ \  $ \\
23.15	&	-13.1	&	1.3	$ \ \ \ \ \  $ \\
28.15	&	-40.9	&	1.4	$ \ \ \ \ \  $ \\
33.15	&	-86.2	&	2.0	$ \ \ \ \ \  $ \\
38.15	&	-144.0	&	2.5	$ \ \ \ \ \  $ \\
43.15	&	-207.9	&	2.4	$ \ \ \ \ \  $ \\
48.15	&	-271.9	&	2.0	$ \ \ \ \ \  $ \\
53.15	&	-331.7	&	1.5	$ \ \ \ \ \  $ \\
58.15	&	-383.7	&	1.2	$ \ \ \ \ \  $ \\
63.15	&	-425.5	&	1.3	$ \ \ \ \ \  $ \\
68.15	&	-455.5	&	1.5	$ \ \ \ \ \  $ \\
73.15	&	-472.6	&	1.6	$ \ \ \ \ \  $ \\
78.15	&	-476.2	&	1.6	$ \ \ \ \ \  $ \\
83.15	&	-466.5	&	1.5	$ \ \ \ \ \  $ \\
88.15	&	-443.8	&	1.3	$ \ \ \ \ \  $ \\
93.15	&	-409.0	&	1.2	$ \ \ \ \ \  $ \\
98.15	&	-363.0	&	1.1	$ \ \ \ \ \  $ \\
103.15	&	-306.9	&	1.1	$ \ \ \ \ \  $ \\
108.15	&	-242.1	&	1.2	$ \ \ \ \ \  $ \\
113.15	&	-169.6	&	1.3	$ \ \ \ \ \  $ \\
118.15	&	-90.8	&	1.3	$ \ \ \ \ \  $ \\
123.15	&	-6.8	&	1.4	$ \ \ \ \ \  $ \\
128.15	&	81.4	&	1.3	$ \ \ \ \ \  $ \\
133.15	&	172.8	&	1.2	$ \ \ \ \ \  $ \\
138.15	&	266.4	&	1.1	$ \ \ \ \ \  $ \\
143.15	&	361.7	&	1.0	$ \ \ \ \ \  $ \\
148.15	&	457.7	&	0.9	$ \ \ \ \ \  $ \\
153.15	&	554.0	&	0.8$ \ \ \ \ \  $ \\	
158.15	&	650.1	&	0.8$ \ \ \ \ \  $ \\	
163.15	&	745.5	&	0.9$ \ \ \ \ \  $ \\	
168.15	&	839.9	&	0.9$ \ \ \ \ \  $ \\	
173.15	&	932.9	&	1.1$ \ \ \ \ \  $ \\	
178.15	&	1024.4	&	1.2$ \ \ \ \ \  $ \\	
183.15	&	1114.1	&	1.3$ \ \ \ \ \  $ \\	
188.15	&	1201.9	&	1.3$ \ \ \ \ \  $ \\	
193.15	&	1287.6	&	1.4$ \ \ \ \ \  $ \\	
198.15	&	1371.2	&	1.4$ \ \ \ \ \  $ \\	
203.15	&	1452.6	&	1.3$ \ \ \ \ \  $ \\	
208.15	&	1531.7	&	1.3$ \ \ \ \ \  $ \\	
213.15	&	1608.6	&	1.2$ \ \ \ \ \  $ \\	
218.15	&	1683.3	&	1.1$ \ \ \ \ \  $ \\	
223.15	&	1755.7	&	1.1$ \ \ \ \ \  $ \\	
228.15	&	1825.9	&	1.0$ \ \ \ \ \  $ \\	
233.15	&	1893.9	&	1.1$ \ \ \ \ \  $ \\	
238.15	&	1959.7	&	1.2$ \ \ \ \ \  $ \\	
243.15	&	2023.5	&	1.3$ \ \ \ \ \  $ \\	
248.15	&	2085.1	&	1.5$ \ \ \ \ \  $ \\	
253.15	&	2144.8	&	1.8$ \ \ \ \ \  $ \\	
258.15	&	2202.4	&	2.1$ \ \ \ \ \  $ \\	
263.15	&	2258.2	&	2.4$ \ \ \ \ \  $ \\	
268.15	&	2312.2	&	2.7$ \ \ \ \ \  $ \\	
273.15	&	2364.3	&	3.0$ \ \ \ \ \  $ \\	
278.15	&	2414.7	&	3.2$ \ \ \ \ \  $ \\	
283.15	&	2463.4	&	3.5$ \ \ \ \ \  $ \\	
288.15	&	2510.5	&	4.9$ \ \ \ \ \  $ \\	
293.15	&	2556.1	&	5.6$ \ \ \ \ \  $ \\	

\hline
\end{tabular}
\end{table}

\nocite{*}

\providecommand{\noopsort}[1]{}\providecommand{\singleletter}[1]{#1}%

\end{document}